\documentclass[12pt]{iopart}
\usepackage{graphicx}
\usepackage{xcolor}

\begin{document}

\title[]{Acoustic spin and  orbital angular momentum  using  evanescent Bessel beams}

\author{Irving Rond\'on}

\address{ School of Computational Sciences, Korea Institute for Advanced Study, \\
Seoul  0245, Republic of Korea.}

\ead{irondon@kias.re.kr}
\vspace{10pt}
%\begin{indented}
%\item[]May 2021
%\end{indented}

\begin{abstract}
Using acoustic evanescent Bessel beam is analyzed the fundamental properties for the spin and orbital angular momentum. The calculation shows that the transversal spin, the canonical momentum, and the orbital angular momentum are proportional to the ratio $l/\omega$,  where  $l$  is the topological charge and $\omega$ the angular frequency.  The complex acoustic Poynting vector and spin density exhibit interesting features related to the electromagnetic case.

\end{abstract}

\section{Introduction}

The wave propagation spin using an evanescent wave has been extensively explored \cite{Bliokh0,Bliokh1,Bliokh2,Aiello,YLong1}; currently, there are many applications where the manipulation of light have an important role, such as topological photonics \cite{Tomoki}, chiral quantum optics \cite{Peter}. Most of these works stated the relationship between the spin angular momentum and its orbital angular momentum, which is ubiquitous, even using intense electromagnetic fields \cite{YiqiFang}.\\
Recently, the acoustic spin and the orbital angular momentum propagation have been reported theoretically \cite{YLong} and experimentally \cite{Shi} using plane waves, for nonparaxial Bessel beam \cite{Bliokh3}. A complete review of the theoretical foundations for the acoustic and electromagnetic field theory has also been reported \cite{Bliokh4}. Applications in acoustic plasmon propagation in waveguides making analogies with electromagnetic fields \cite{DanielLeykam}, acoustic surface waves \cite{Transv}, acoustic radiation forces \cite{Marston,MarstonZhang,LZhang,Toftul,Glauber}, enhanced manipulation in near fields \cite{Gires}, in fluids physics propagating vortex  \cite{RondonLeykam}, in inhomogeneous media \cite{Fan} using meta-surface  waveguides have been demonstrated applications on spin-related robust transport \cite{YLong2}, and the importance to  explore symmetry properties to enhanced acoustic propagation in the near field \cite{YLong3}.\\
\\
The evanescent acoustic waves have been used in several biomedical engineering applications. For example, some research groups have developed methods for the flexible trapping and spinning of individual living cells, among others see, e.g. \cite{MMarte} and references therein.  An overview of the analytical strategies that employ evanescent-wave optical biosensors to deal with the complexities and challenges of effective Nucleic Acids, DNA, and RNA detection was presented by \cite{CHuertas}.\\
In this direction, our contribution in this work is to study evanescent wave following the approach presented in \cite{Bliokh3,Bliokh4}, which allows describing the arbitrary acoustic field and deriving quantities such as the time average energy density, the Poynting vector, the canonical momentum, the spin, and orbital angular momentum. A qualitative  description for the dynamical quantities between the acoustic field and the electromagnetic case \cite{LDu,Tsesses}; The results shows that an evanescent wave, the traversal acoustic spin vector and canonical momentum are proportional to the factor $l/\omega$, as predicted in acoustics vortex propagation in free space \cite{Marston,MarstonZhang,LZhang,Karen}.\\
In sections 2 and 3, the basic fundamental equations describe the spin acoustic wave propagation and its fundamental properties. The results follow the principal features and the physical propagation quantities for acoustic evanescent Bessel beam, and the final remarks are presented in section 4. 

\section{Basic fundamental background theory for acoustic beam}
The fundamental equations for fluids physics in the absence of external forces in the linear  regime are \cite{Landau,Bruneau},

\begin{equation}
\label{Eq1}	
\rho	\frac{\partial  \vec{v}}{\partial t} = -\nabla P ,	
\end{equation}

\begin{equation}
	\label{Eq2}
	\beta \frac{\partial P}{\partial t} = -\nabla \cdot \vec{v},	
\end{equation}
these equations express basic  medium parameters such as $\rho$ mass density, $\beta=1/V$, the bulk modulus. For monochromatic acoustic  waves the pressure and  velocity are expressed as $P(\vec{r},t) = \mathbf{Re}[P(\vec{r})\, e^{-i\omega t}]$ ,  $\vec{v}(\vec{r},t)= \mathbf{Re}[\vec{v} (\vec{r})\,  e^{-i\omega t}]$. 
 Substituting them into  Eqs  . \ref{Eq1} and \ref{Eq2},  the dynamical equations read as
\begin{equation}
		\label{Eq3}
	\vec{v}=   \frac{1}{i \rho \omega}\, \nabla P,
\end{equation}
and
\begin{equation}
	\label{Eq4}
	P=  \frac{1}{i \beta \omega} \,  \nabla \cdot \vec{v}.
\end{equation}
It is important to mention that on taking the rotational of Eq. \ref{Eq3}, there  results in a curl-free acoustic filed velocity, that is $\nabla \times \vec{v}=0$ \cite{Landau,Bruneau} .  Other interesting features are that Eqs. \ref{Eq3} and \ref{Eq4} can be related by the plane angular spectrum \cite{Whittaker}. This representation is very well known in optics, and it allows the study and design of different propagating beams; for a review, see \cite{Uri} and references therein. 
The following section is presented the fundamental properties for an evanescent Bessel beam using the recent approach presented in \cite{Bliokh3,Bliokh4,YLong3}
\begin{figure} 
	\centering
	\includegraphics[scale=1.20]{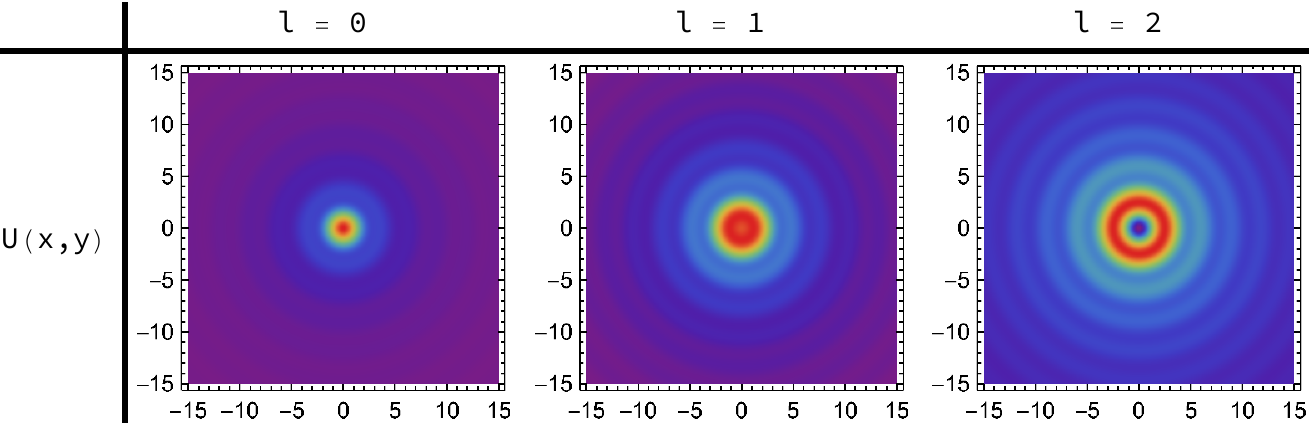} 
	\caption{ The total transversal energy Eq. \ref{eqTotalEnergy} using different topological charges.} 
	\label{Figure1}	
\end{figure}

\section{Evanescent acoustic Bessel beam} 
In this work,  the study for  sound waves  represented for  evanescent  Bessel beams is given by  

\begin{equation}
	\label{EqPressure}
	P= P_0 J_l(\mu r ) e^{ - \kappa z + i l \theta    },
\end{equation}
where $P_0$ is given amplitude (initial pressure), $l$ the topological charge, the  wave vector dispersion relation for a evanescent wave $\mu = \sqrt{k^2 + \kappa^2}$, where the wave vector is given by  $k=\omega \sqrt{\rho \beta}$ and $r=\sqrt{x^2 +y^2}$. Note that the wave vector magnitude $\mu$ follows the same notation reported  in \cite{Yang}.\\
Using the scalar pressure for the  evanescent  Bessel beam given by  Eq. \ref{EqPressure}, it is derived the acoustic vector velocity field  using Eq. \ref{Eq3}   to obtain 
\begin{equation}
\label{EqVelocity}
\vec{v}=v_0 
\big[-i \mu  (J_{l -1}(\mu r)-J_{l +1}(\mu r)),\mu  (J_{l -1}(\mu  r)+J_{l +1}(\mu r)),2 i  \kappa J_{l }(\mu r) \big]e^{i l \theta - \kappa z  } ,	
\end{equation}
where $v_0= P_0/ 2\omega \rho$,   and it  used the cylindrical-coordinate components. Using Eqs. \ref{EqPressure}  and \ref{EqVelocity} to describe acoustic evanescent Bessel beams, the fundamental propagation properties  will be the following:
\begin{figure} 
	\centering
	\includegraphics[scale=0.5]{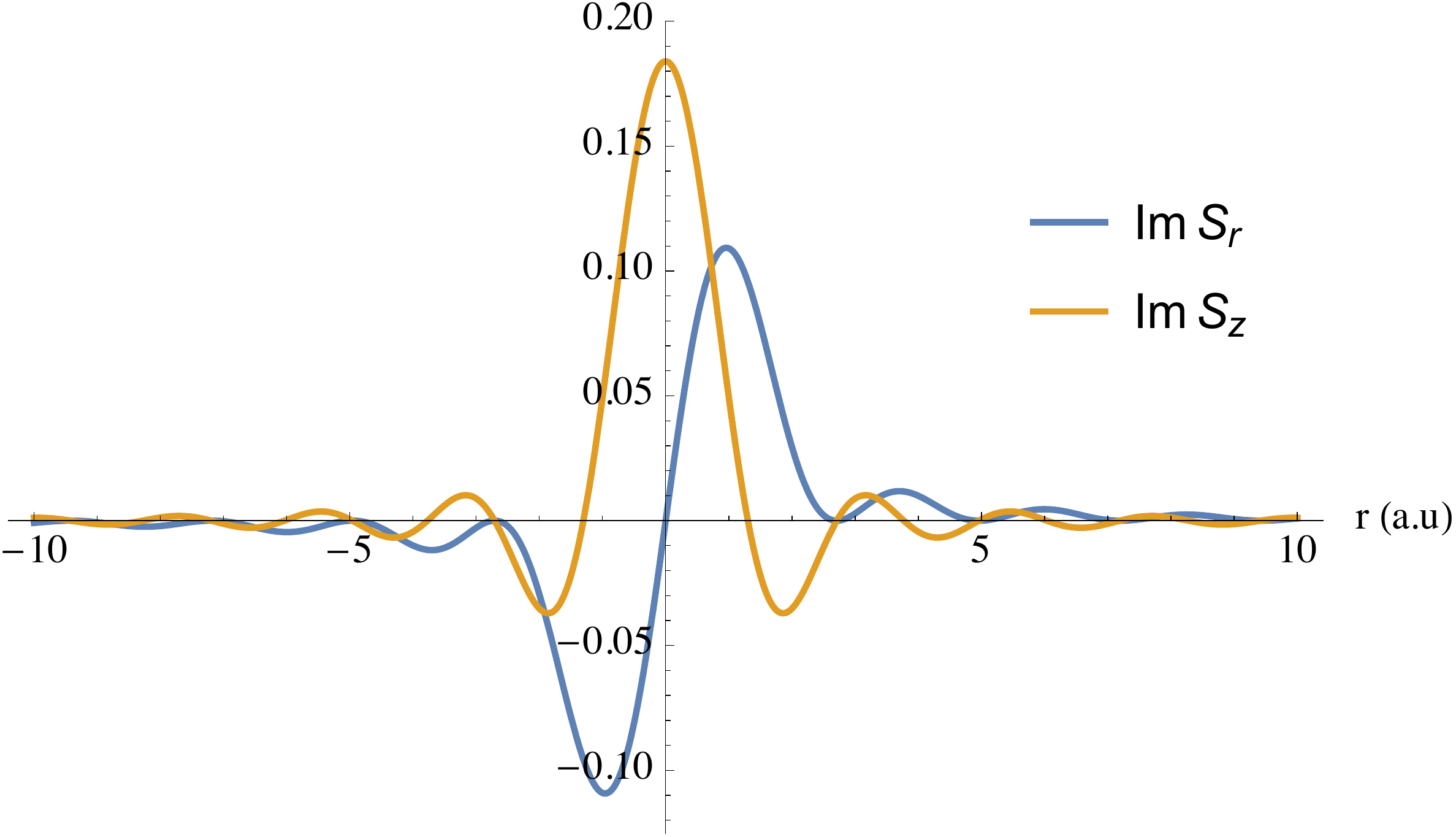} 
	\includegraphics[scale=0.5]{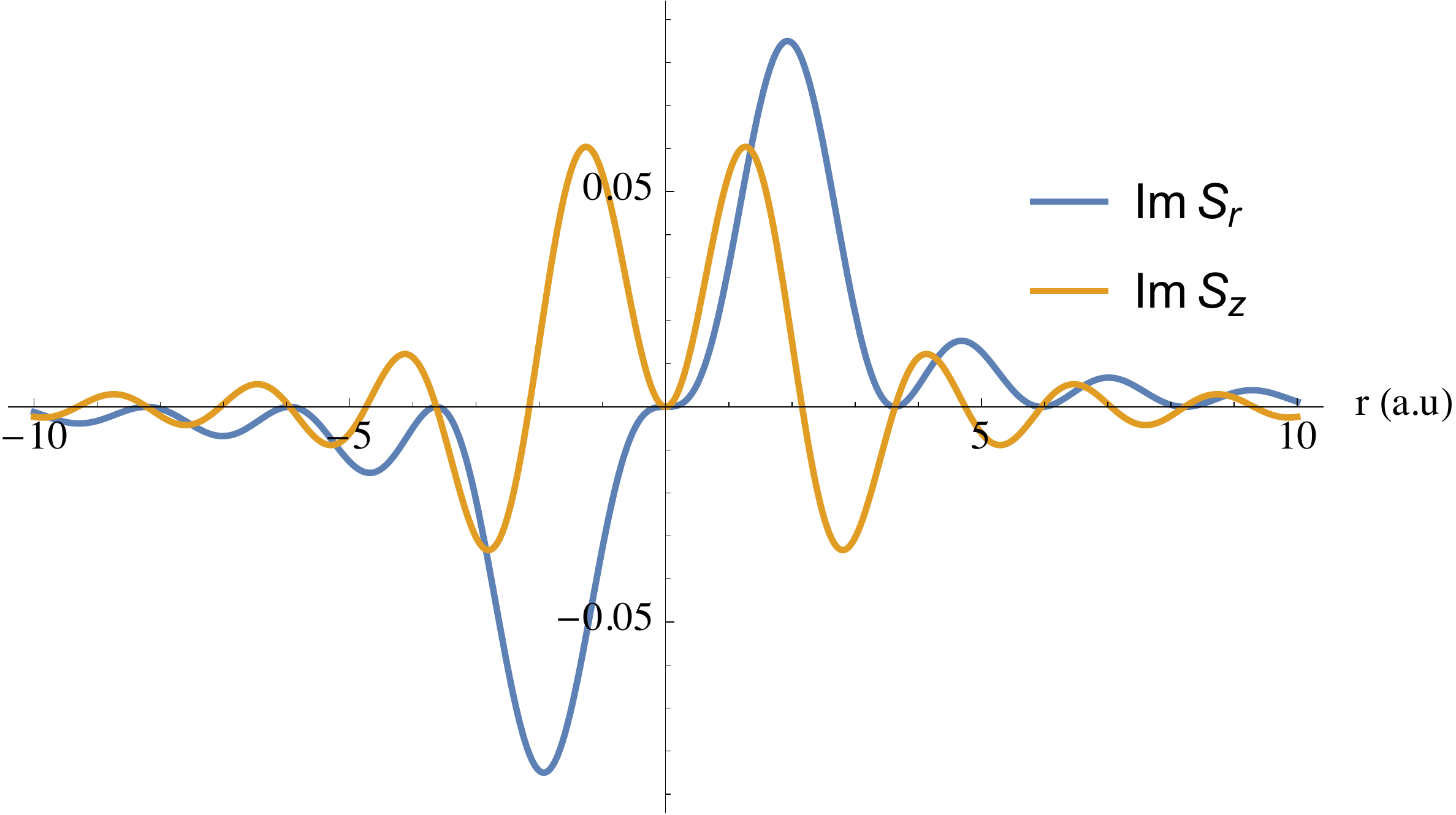} 
	\caption{The longitudinal and radial spin vector component Eq. \ref{Spin}  with different topological charges $l=1$ (upper) and $l=2$  (bottom). Similar behavior for spin density in the optical case was reported in  Ref. \cite{LDu}. } 
	\label{Figure2}
\end{figure}

\subsection{The acoustic energy density acoustic}
\noindent
The time average  acoustic  energy density is  
\begin{equation}
	\label{EqW}
U = \frac{1}{4} \left( \beta  |P|^2 + \rho |\vec{v}|^2 \right),	
\end{equation}
after calculations, it is possible to  write the energy density in term of  the local pressure and velocity as   
\begin{equation}
	U = 
	U_P + U_v, 
\end{equation}
 After  calculation, the energy density distribution is 

\begin{eqnarray}
		\label{eqWp}	
	U_P& =  u_0  \,  J_l^2( \mu r ) e^{-2 \kappa  z} ,\\
	U_v &=  u_0 \left(\frac{\kappa^2}{k^2} J_{l}^2( \mu r )+  \frac{\mu^2}{2 k^2} ( J_{l-1}^2( \mu r )+ J_{l+1}^2( \mu r ))\right)  e^{-2 \kappa  z},
	\label{eqWv}
\end{eqnarray}
for simplicity, we define the factor $u_0= \beta P_0^2/4$,
\begin{equation}
	\label{totalU}
	U =  
	u_0  \left(  ( 1  + \frac{\kappa^2}{k^2} )J_{l}^2( \mu r )+  \frac{\mu^2}{2 k^2} ( J_{l-1}^2( \mu r )+ J_{l+1}^2( \mu r ))\right)  e^{-2 \kappa  z};
\end{equation}
moreover, it is possible to recast the quantities using  wave vector dispersion relation, and obtain an approximate factor as $\mu^2/k^2= 1+ \kappa^2/k^2$. Using this consideration, the total acoustic density energy can be written as
\begin{equation}
	\label{eqTotalEnergy}	
	U=  \frac{ \mu ^2}{ k^2}  \left(   J_l^2( \mu r )+ \frac{1}{2} \left[J_{l-1}^2( \mu r )+J_{l+1}^2( \mu r )\right]\right) u_0 e^{-2 \kappa  z},
\end{equation}
It is important to remark that a similar factor was reported in optical evanescent Bessel  \cite{Yang}.  Experimentally this kind of correction  \cite{ZYang1} can be achieved using spiral phase plate.  In particular, when $U(0)$ has a nonzero energy density in the center as it is  shown in  Fig. \ref{Figure1}. 

\subsection{The acoustic spin density}
Using  the scalar pressure  Eq. \ref{EqPressure}, it is possible to generate a  vector velocity field distribution Eq. \ref{EqVelocity},  it allows to  calculate  the acoustic spin density   using the following relation \cite{Bliokh3} as
\begin{equation}
\vec{S}= \frac{\rho}{2 \omega} \mathbf{Im}(\vec{v^{*}}\times \vec{v}),	
\end{equation}
where $\rho$  is the mass density, $\omega$  is the frequency, and $\vec{v}$ is the acoustic velocity field given by Eq. \ref{EqVelocity}, 
\begin{equation}
	\label{Spin}	
\vec{S}=  S_0 \left(  \frac{ \kappa  l }{  r }  J_l^2( \mu r ) 
\hat{e}_r +
\frac{ \mu ^2  }{4 }\left[ J_{l-1}^2(\mu r )-J_{l+1}^2(\mu r )\right] \hat{e}_z\right)  e^{-2 \kappa  z},	
\end{equation}
where  $S_0= P_0^2  /\rho \omega ^3 $, this result Eq. \ref{Spin}  represents the acoustic spin, and shows that a sound evanescent Bessel beam has well-defined radial and longitudinal components. \\ 
It is straightforward  to recast  Eq. \ref{Spin} into Cartesian coordinates using,  $S_x = S_r \cos\theta  - S_\theta \sin \theta$ and $S_y= S_r \sin \theta + S_\theta \cos \theta$, where the angular component is $S_\theta=0$,  the acoustic spin results as

\begin{equation}
\label{EqSx}
	S_x =  S_0 \frac{ \kappa  l }{  r }  J_l^2( \mu r ) \cos \theta \,  e^{-2 \kappa  z},
\end{equation}
 
 \begin{equation}
 \label{EqSy}
 	S_y =  S_0 \frac{ \kappa  l }{  r }  J_l^2( \mu r ) \sin \theta  \, e^{-2 \kappa  z} , 
 \end{equation}
 and
  \begin{equation}
  \label{EqSz}
 	S_z =  S_0 \frac{ \mu ^2  }{4 }\left( J_{l-1}^2(\mu r )-J_{l+1}^2(\mu r )\right)  e^{-2 \kappa  z}.
 \end{equation} 
These results show that for $l=0$ non-vortex is obtained, and for $l \ne 0$, the acoustic spin has radial and longitudinal components.  
\\
It is possible to study the radii between the total  energy Eq.  \ref{totalU} and Eq. \ref{Spin}  to obtain the following relations
\begin{equation}
\frac{ S_r}{U}  = \frac{l}{\omega } \frac{J_l^2(\mu r )}{r} \frac{ 4 \kappa /  \mu^2}{ \left(J_l^2( \mu r ) +\frac{1}{2} \left(J_{l-1}^2( \mu r )+J_{l+1}^2(\mu r)\right)\right)}, 
\end{equation}
and 
\begin{equation}
	\frac{S_z}{U} = \frac{1}{\omega} \, \frac{J_{l-1}^2(\mu r )-J_{l+1}^2(\mu r )  }{ \left( J_l^2( \mu r )+ \frac{1}{2} \left(J_{l-1}^2( \mu r )+J_{l+1}^2( \mu r )\right). \right)}.	
\end{equation}
These expression show that the radial component is proportional to beam topological charge and angular frequency 
$l/\omega$ \cite{Karen}. The expression $ S_r/U$  is dependent of  the  $l$ sign. The longitudinal radii $S_z/U$   is related  as $l-1$ and $l+1$. Substituting the value $l=0$ into  Eqs. \ref{EqSx},\ref{EqSy} and \ref{EqSz} the radial and longitudinal component vanish. For this value the fields does not interact in such a way to generate any interference. The   following Bessel function property  $J_{-l}(x ) = (-1)^l J_{l}(x) $ can be used. In Fig. \ref{Figure2}  correspond numerically the radial variation of the transverse $S_r$ and longitudinal $S_z$ spin component of the beam using different order $l$.\\
\begin{figure} 
	\centering
	\includegraphics[scale=1.2]{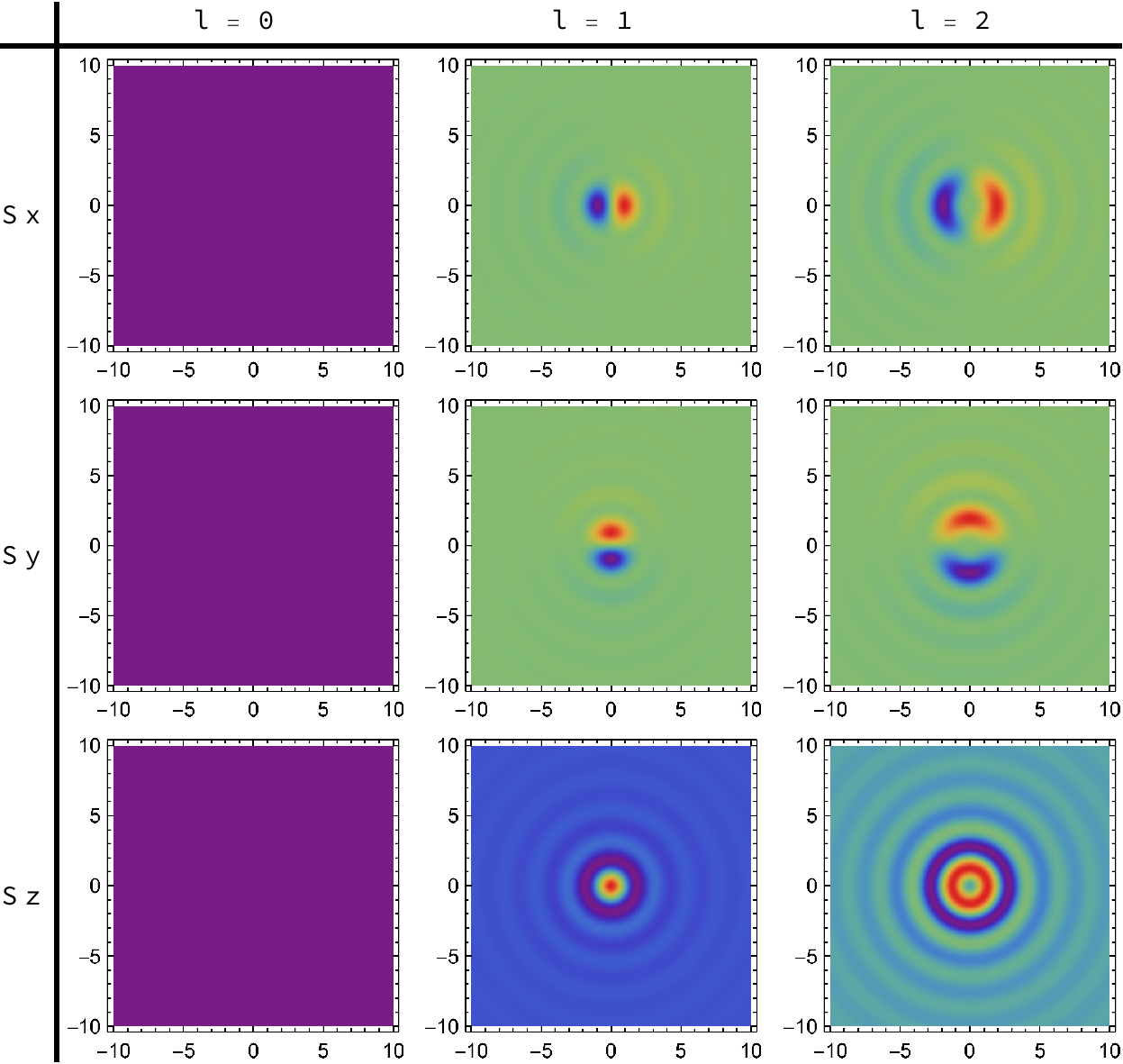} 
	\caption{  Numerical behavior  for the acoustic spin density  Eq.
		\ref{EqSx} , \ref{EqSy} and  \ref{EqSz} in Cartesian coordinates using different  $l$ values.   
		\\ } 
	\label{Figure3}
\end{figure}
In Fig \ref{Figure3} are  shown numerically the acoustic spin density, it is transversal out of  a plane and longitudinal oriented spin, $S_x$ exhibit two-lobe pattern along $x$ axis, for $S_y$ are arranged along $y$ axis, this two distribution $S_x$ and $S_y$ are rotated by $90^{\circ}$  relative to each other. For $S_z$ a well-defined acoustic spin density, the maximum spin value is obtained for $l=1$ at $x=0$, and it is divided in two by increasing the topological charge $l=2$. For $l=0$,  all acoustic spin components are zero since these quantities are indeterminate for $r=0$.\\
Furthermore, these results resemble the case studied in \cite{LDu} using an electromagnetic evanescent Bessel beam. In this case, the optical spin has radial and longitudinal components. For instance, making analogies with Eq.\ref{Spin}, it is possible to predict similar behavior experimentally using an evanescent acoustic Bessel beam. 
\\
Finally, using the acoustic density energy Eq.  \ref{eqWp} allows writing a transversal spin vector using the Eqs. \ref{EqSx} and \ref{EqSy}  as
\begin{equation}
	\label{eqSxSy}	
	(S_x,S_y) = \frac{l}{\omega} \gamma  \left( \frac{\cos\theta}{r}  , \frac{\sin \theta  }{r} \right) U_P  
\end{equation}
where $\gamma= 4 \kappa /k^2$, taking $ \vec{u}= \frac{\gamma}{r}  \left( \cos\theta  , \sin \theta \right)$,  this  expression can be written as
\begin{equation}
	\label{StUp}
	\vec{S_t}= \frac{l}{\omega} \, U_P\, \vec{u},
\end{equation}
this result shows that the evanescent acoustics spin vector is proportional to $l/\omega$.

\subsection{The acoustic Poynting vector}
In Refs. \cite{YLong2,YLong3}  the authors have shown the vector propagation of the reactive power and time average flow in the Poynting vector to understand the acoustics surface wave in meta-surfaces. These properties are calculated using an evanescent acoustic beam.
\\
The time average energy is  given by \cite{Bliokh3,Bliokh4} as: 
\begin{equation}
\label{VecAPoynt}	
		\vec{\Pi}=  \frac{1}{2}\mathbf{Re}\left[P^* \vec{v}\right],
\end{equation}
and the reactive component  using \cite{Yang}
\begin{equation}
\label{VecRPoynt}
	\vec{R}=  \frac{1}{2}\mathbf{Im}\left[P^* \vec{v}\right],	
\end{equation}
using Eq. \ref{EqPressure} and  \ref{EqVelocity} is straightforward to  obtain the time average energy flow
\begin{equation}
		\Pi_\theta=  \frac{P_0^2 \mu}{4 \rho  \omega }  J_l(r \mu ) (J_{l-1}(r \mu )+J_{l+1}(r \mu )) e^{-2 \kappa  z},
\end{equation}
the result shows that the time average  energy flow only has an angular component described by a Bessel function of order $l-1$ and $l+1$, typically related to circularly polarized basis, which carries orbital angular momentum. \\
For the reactive acoustic Poynting vector as
\begin{equation}
		\vec{R} = \frac{1}{4 \rho  \omega} \left(   \mu  J_l (r \mu) \left( J_{l+1}\left(r \mu \right)  -J_{l-1} (r \mu )\right)\hat{e}_r +	2\kappa  J_l^2 (r \mu  )\hat{e}_z \right)  P_0^2  e^{-2 \kappa  z},
\end{equation}
the  reactive vector has radial and longitudinal components.   Note that the spin angular momentum  is orthogonal to the energy flux density $\vec{\Pi} \cdot \vec{S}=0$ and  $\vec{R} \cdot \vec{S}=0$. 

\begin{figure} 
	%\centering
	\includegraphics[scale=0.7]{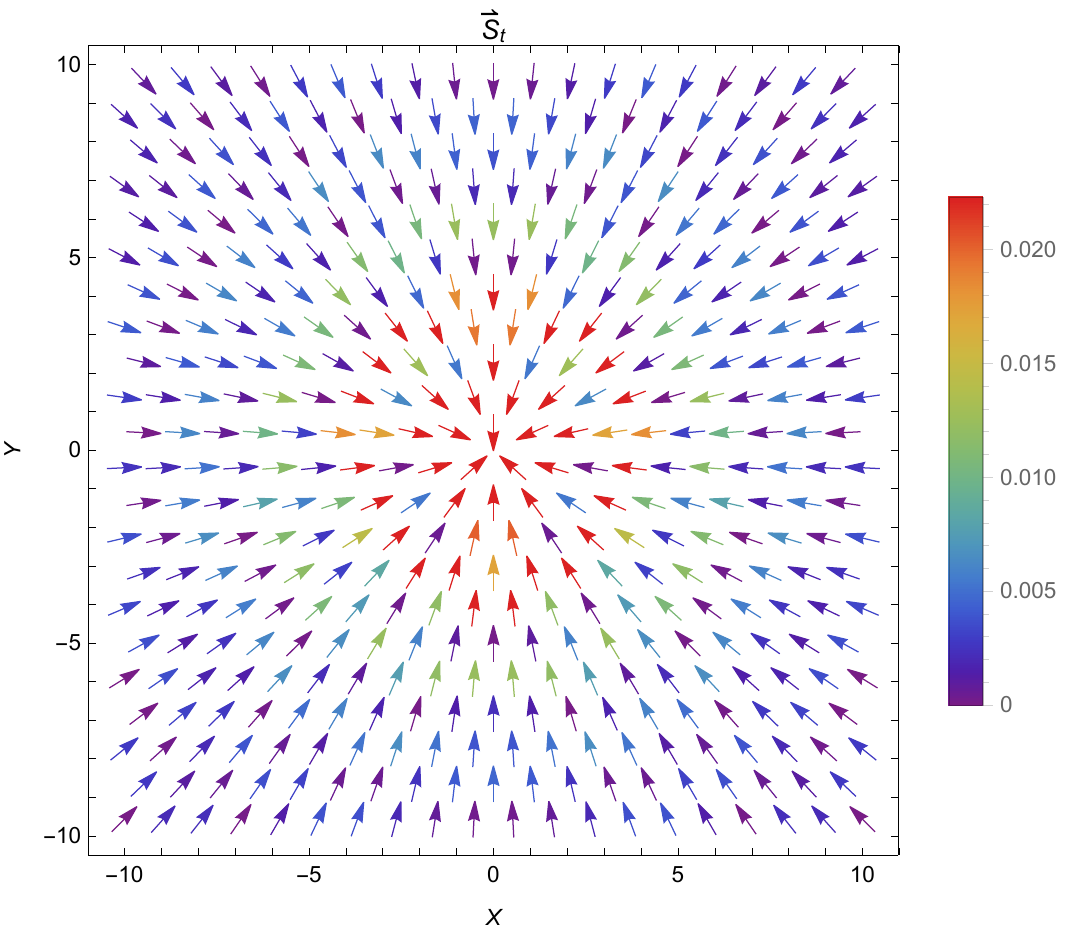} 
	\includegraphics[scale=0.7]{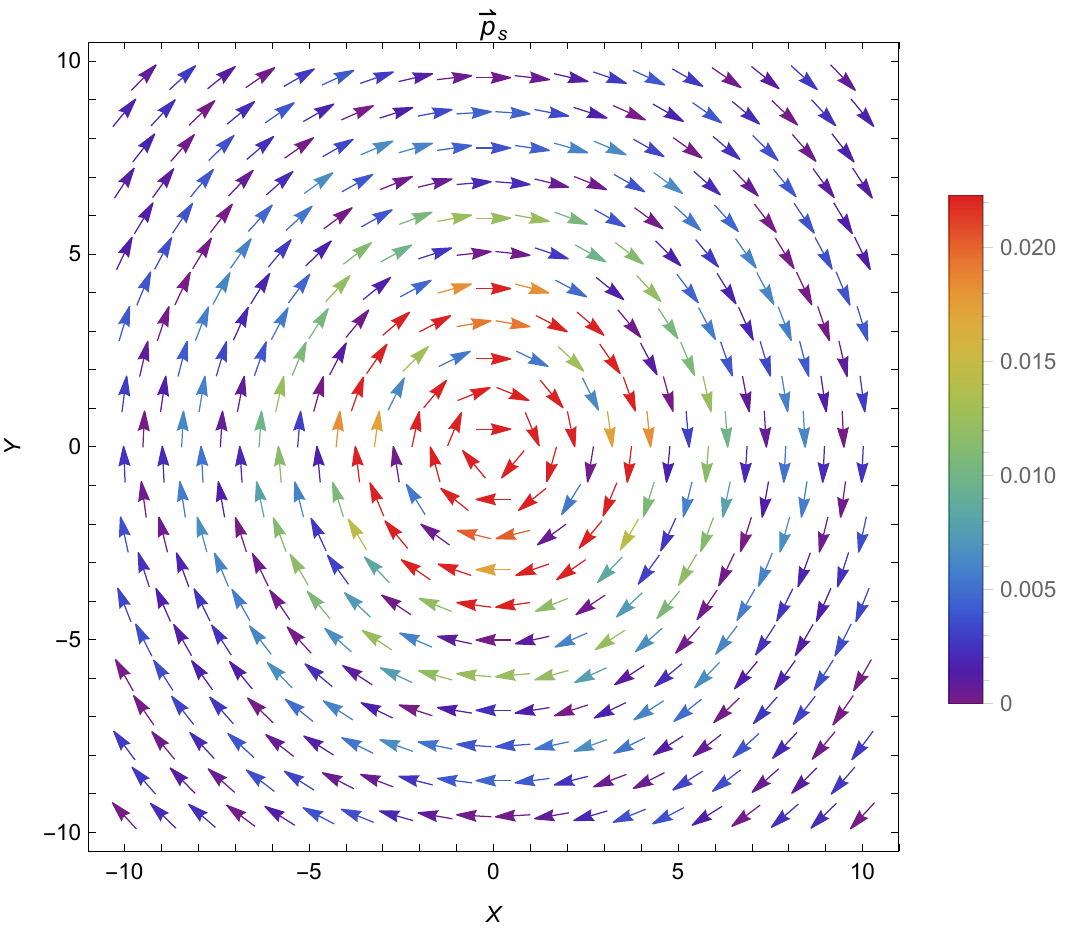} 
A qualitative behavior for transversal vector acoustic spin Eqs. \ref{StUp}   (left panel) for the transversal spin momentum density Eq. \ref{ptUp} (right panel),  in both cases circular behavior is presented and a  value of $l=-1$ was taken. Note that different sign values of $l$ can show an inward/outward or change in the rotation spin.
	\label{Figure4}
\end{figure} 

\subsection{The canonical momentum}
The authors have shown that the canonical momentum   \cite{Bliokh3,Bliokh4}  can be expressed as 
\begin{equation}
	\vec{\Pi} = 	\vec{p}_{\,o} + 	\vec{p}_{\,s},
\end{equation}
where $\vec{p}_{\,o}$ and $\vec{p}_{\,s} $ are given by

\begin{equation}
\label{Porb}	
	\vec{p}_{o} = \frac{\rho}{2 \omega}  \mathbf{Im} \left[  \vec{v^*} \cdot (\nabla)\, \vec{v}\right],
\end{equation}
\begin{equation}
\label{Pspin}	
	\vec{p}_{\,s} = \frac{\rho}{4 \omega}  \left[ \nabla \times \mathbf{Im} ( \vec{v^{*}} \times \vec{v})\right] .
\end{equation}
$\vec{\Pi }$ is the Poynting vector Eq.  \ref{VecAPoynt} , it represents  the kinematic momentum (or Poynting momentum)  $\vec{p}_{\,o}$ is the  orbital momentum Eq. \ref{Porb}
and their difference gives  $	\vec{p}_{\,s}$ the spin momentum Eq. \ref{Pspin}. For  more details see  \cite{Bliokh3,Bliokh4} and references therein.
\\
Using  Eq. \ref{Porb},   the  orbital  momentum density  is
\begin{equation}
	\label{Porb_t}
	\vec{p}_{o}=  \hat{e_\theta}\frac{ l}{\omega } \frac{u_0}{ r}  	\left(\frac{\kappa ^2 }{k^2} J^2_l( \mu r ) +
	\frac{\mu ^2}{ 4 k^2} (J_{l-1}(r \mu )-J_{l+1}( \mu r ))^2\right) e^{-2 \kappa  z},	
\end{equation}
for $l=0$ the orbital momentum is zero, when $l > 0$ this expression is well defined. 
\\
The spin density Eq. \ref{Pspin} gives 
\begin{equation}
		\label{Pspin_t}
\vec{p}_{s} = \hat{e}_\theta \frac{ l}{\omega  } \,  \frac{J^2_l(r \mu )}{r}  u_0 e^{-2 \kappa  z},
\end{equation}
this result shows
that the transversal acoustic momentum for evanescent is proportional to  $l/\omega$, and have a purely angular dependence along the propagation axis. It is well defined except for $r=0$.
After transforming  Eq. \ref{Pspin} into  Cartesian coordinates using  $p_x = p_r \cos\theta  - p_\theta \sin \theta$, $p_y= p_r \sin \theta + p_\theta \cos \theta$, as $p_r=0$,  and taking $ \vec{u'}= \frac{1}{r}  \left( -\sin\theta  , \cos \theta \right)$, with the density energy Eq.  \ref{eqWp}, it allows  to write the acoustic spin vector as   
\begin{equation}
	\label{ptUp}
\vec{p}_{s}=  \frac{l}{\omega} \, U_P\, \vec{u^{'}} .	
\end{equation}
Here it is important to remark that 
our results given by Eq. \ref{StUp} and Eq. \ref{ptUp} can be related to previous results, where the authors have been demonstrated that the torque can be related with the power absorbed as  $T = \left(  l/\omega  \right) P_{abs}$ using an acoustic Bessel beams  \cite{Marston,LZhang}.   In Fig 4, it is shown the qualitatively vector behavior for the acoustical spin Eq. \ref{StUp}   and momentum spin density Eq. \ref{ptUp}.

\begin{figure} 
	\centering
	\includegraphics[scale=1.3]{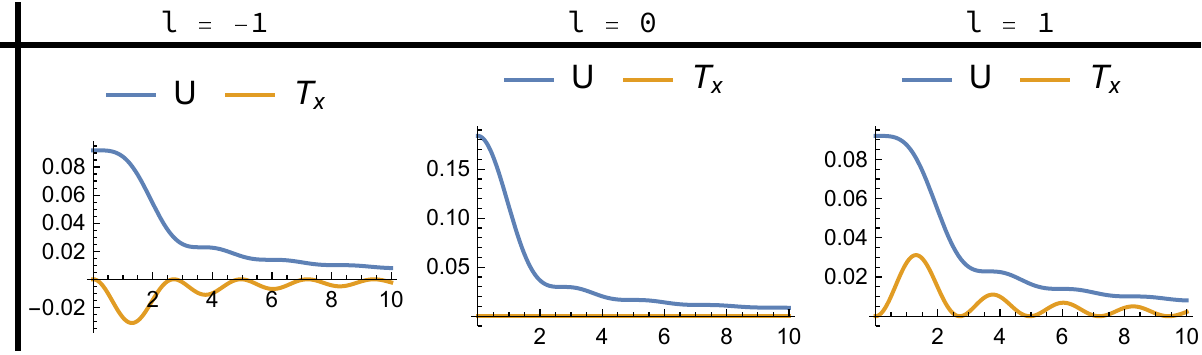} 
\caption{The transversal torque  Eq. \ref{Torque}, shows that its local behavior is dependent on the sign value of $l$, with the total acoustic energy density  Eq. \ref{eqTotalEnergy}. Similar behavior has been reported in Ref. \cite{LZhang} using acoustic Bessel beams.} 
	\label{Figure5}
\end{figure}

\subsection{The orbital angular momentum and torque}
\noindent
For  acoustic beam the orbital angular momentum is 
\begin{equation}
	\vec{L}= \vec{r}\times\vec{p} = \vec{r}\times(\vec{p}_{\,s}+ \vec{p}_{\,o}),
\end{equation}
physically  the only contribution is given by  orbital  momentum   \cite{YLong3}

\begin{equation}
		\vec{L}= \vec{r}\times\vec{p}_{o} ,
\end{equation}
after a straightforward calculation, it is shown that the orbital angular momentum propagation axis $z$ is

\begin{equation}
\label{Lz}
L_z= \frac{l}{\omega }    \left(  \frac{\kappa^2}{k^2} J^2_l(r\mu)+ \frac{\mu ^2}{4 k^2}
(J_{l-1}(r \mu )-J_{l+1}(r \mu ))^2 \right) u_0
e^{-2 \kappa  z}, 	
\end{equation}
this result shows that the orbital angular momentum (OAM) of evanescent acoustic waves is well defined and proportional to  $l/\omega$  \cite{Marston,LZhang,Karen,ZYang1}. Another interesting result is that the radii $L_z/U$ also present the factor $(1+ \kappa^2/k^2)$ reported in optical evanescent  Bessel beam \cite{Yang}.  Similarly, as the acoustic Bessel beam case using  Eqs.  \ref{Porb} and  \ref{Lz}  this result shows that
the orbital angular momentum and the canonical momentum are related by $ L_z= r \, p^{o}_\theta$. Details account for the orbital angular momentum and properties described
bellow for acoustic beams can be found in \cite{Bliokh3,Bliokh4,YLong3}.\\
Finally, it is  derived the acoustic torque on a small absorbing isotropic dielectric particle immersed in a monochromatic field, following  the expression  \cite{Toftul}.
\begin{equation}
	\vec{T} = \frac{1}{2} \mathbf{Re}[\rho \vec{D}^{*}\times \vec{v}]= \omega\,  \mathbf{Im}(\alpha_d)\vec{S},	
\end{equation}
where $\vec{D}=\alpha_d \vec{v}$,  represents the  dipole moment, and $\alpha_d$ is the complex dipole polarizability of the particle, using  Eq. \ref{Spin}, the transversal torque is

\begin{equation}
	T_r = \omega\,  \mathbf{Im}(\alpha_d)
	S_0 \frac{ \kappa  l }{  r }  J_l^2( \mu r ) 
	e^{-2 \kappa  z},
\end{equation}
after transforming in Cartesian coordinates and using  the result expressed in Eq. \ref{StUp}  an expression for transversal  acoustic spin is  obtained
\begin{equation}
	\label{Torque}
(T_x,T_y)= l   \, U_P\, \vec{u},
\end{equation}
and the longitudinal component  
\begin{equation}
	T_z= \omega\,  \mathbf{Im}(\alpha_d) 
	\frac{ \mu ^2  S_0 }{4 }\left(J_{l-1}^2(\mu r )-J_{l+1}^2(\mu r )\right)   e^{-2 \kappa  z},
\end{equation}
In figure 5 is shown the numerical behavior of Eq. \ref{Torque}, the uni-dimensional plot of $T_x$ behavior for different $l$ values.
These results  can be extended for any arbitrary acoustic beams, 
and they agree with previous studies \cite{Bliokh3, Bliokh4,LZhang,Toftul,Karen}. Further analysis is beyond the scope of this work, but it can open further exploration in order to find new applications such as acoustic trapping, surface waves, and acoustic forces. One interesting direction is using the shape beams approach used in the  Mie scattering approximation \cite{Gouesbet}. 

\section{Summary}
This work presented fundamental local properties for the acoustic field:  energy, canonical momentum, spin angular momentum densities, and the complex acoustic  Poynting vector for evanescent Bessel beam,  which resulted in an agreement. This approach can be applied to any arbitrary acoustic beams. It is important to note that the energy density pressure is directly related to the angular spectrum representation. 
Other pressure-structured fields related to the angular spectrum
representation commonly used in optics to generated Xwaves, Frozen waves among others structured beams can be used \cite{Hugo2014}.
These results show that the transversal acoustic spin and the canonical momentum vector proportional to $l/\omega$ and the transversal torque as  $\vec{T}_t = l \, U_P \vec{u}$,   reported in acoustic force analysis \cite{Marston,MarstonZhang,LZhang}. 
The main idea of this work was to show the possible connections among optics, electromagnetic and sound waves and where the acoustic spin density can bring interest by studying the remarked similarities among evanescent waves  \cite{MMarte,CHuertas,LDu,Tsesses}.
Using this kind of analogies may be helpful to explore new possibilities in the acoustic control of the direction of motion and speed of nano-objects in a layer of biological fluid as the red blood cells
\cite{OlegV1,OlegV2}.

%%%%%%%%%%%%%%%%%%%%%%%%%%%%%%%%%%%%%%%%%%

\section{Acknowledgments}
{\noindent
	This research was   supported by Basic Science Research Program through the National Research Foundation of Korea (NRF) funded by the Ministry of  Science and ICT [NRF-2017R1E1A1A01077717.}

%=====================================


\section{References}
\begin{thebibliography}{99}
	
	\bibitem{Bliokh0} K. Y. Bliokh, A. Y.Bekshaev, F. Nori 2014 Extraordinary momentum and spin in evanescent waves. Nat. Commun., 5, 3300.
	
	\bibitem{Bliokh1}   K. Y. Bliokh, F Nori 2015 Transverse and longitudinal angular momenta of light, Phys. Rep. 592, 1–38 .	
	
	\bibitem{Bliokh2} K. Y. Bliokh, F. J. Rodríguez-Fortuño, F. Nori, and A.V  Zayats 2015 Spin–orbit interactions of light, Nat. Photon. 9, 796–808 .
	
	
	
	\bibitem{Aiello} A. Aiello, P. Banzer, M. Neugebauer, G Leuchs 2015 From transverse angular momentum to photonic wheels, Nat. Photon. 9, 789–795 .	
	
	\bibitem{YLong1} Y. Long, J.  Ren, H. Chen  2018 Intrinsic spin of elastic waves. Proc. Natl. Acad.Sci. USA 115, 9951–9955.
	
	
	\bibitem{Tomoki} T. Ozawa, et al.  2019 Topological photonics, 
	Rev. Mod. Phys. 91, 015006 .	
	
	
	\bibitem{Peter} P. Lodahl, et al 2017  Chiral quantum optics, Nature 541, 473–480.
	
	\bibitem{YiqiFang} Y. Fang et al  2021 Photoelectronic mapping of the spin–orbit interaction of intense light,
	Nature Photonics 15, 115–120.
	
	\bibitem{YLong}  Y. Long,  J. Ren, and H. Chen 2018 Intrinsic spin of elastic waves
	PNAS 2018 , 115 ,40, 9951-9955.
	
	\bibitem{Shi} C. Shi, et al. Observation of acoustic spin 2019 Natl. Sci. Rev. 6, 707–712.	
	
	\bibitem{Bliokh3}  K. Y. Bliokh, F. Nori, Spin and orbital angular momenta of acoustic beams  2019 Phys. Rev. 99 174310.
	
	\bibitem{Bliokh4}L. Burns, K. Y. Bliokh, F. Nori, J. Dressel 2020 Acoustic versus electromagnetic field theory: scalar, vector, spinor representations and the emergence of acoustic spin, New J. Phys. 22 , 053050.
	
	\bibitem{DanielLeykam}Daniel Leykam, Konstantin Y. Bliokh, and Franco Nori 2020
	Edge modes in two-dimensional electromagnetic slab waveguides: Analogs of acoustic plasmons,Phys. Rev. B 102, 045129.
	
	
	\bibitem{Transv} K. Y. Bliokh, F. Nori, Transverse spin and surface waves in acoustic metamaterials 2019 Phys. Rev. B 99, 020301.
	
	\bibitem{Marston} L. Zhang and P. L. Marston  2011 Angular momentum flux of nonparaxial acoustic vortex beams and torques on axisymmetric objects, Phys. Rev. E 84, 065601(R).
	
	\bibitem{MarstonZhang} L. Zhang, and  P. L. Marston,  2013 Optical theorem for acoustic non-diffracting beams and application to radiation force and torque, Biomed. Opt. Express 4, 1610-1617 .
	
	\bibitem{LZhang} L. Zhang 2018 Reversals of Orbital Angular Momentum Transfer and Radiation Torque, Phys. Rev. Applied 10, 034039 .
	

	
	\bibitem{Toftul} I. Toftul, K. Bliokh, M Petrov, F. Nori  2019  Acoustic radiation force and torque on small particles as measures of the canonical momentum and spin densities, Phys. Rev. Lett. 123, 183901.
	
	\bibitem{Glauber}J. H. Lopes, E. B. Lima, J. P. Leão-Neto, and G. T. Silva 2020 Acoustic spin transfer to a subwavelength spheroidal particle, Phys. Rev. E 101, 043102 .
	
	\bibitem{Gires} Pierre-Yves Gires, C. Poulain  2019 Near-field acoustic manipulation in a confined evanescent Bessel beam. Commun Phys 2, 94 .
	
	
	\bibitem{RondonLeykam} I. Rond\'on and D. Leykam  2020 Acoustic vortex beams in synthetic magnetic fields, J. Phys.: Condens. Matter 32, 104001. 
	
	
	\bibitem{Fan} X.D. Fan, Z. Zou, and L. Zhang  2019 Acoustic vortices in inhomogeneous media, Phys. Rev. Research 1, 032014.
	
	\bibitem{YLong2}Y. Long, et al 2020 Realization of acoustic spin transport in metasurface waveguides. Nat Commun 11, 4716.
	
	
	\bibitem{YLong3} Y. Long et al. Symmetry selective directionality in near-field acoustics 2020 Natl. Sci. Rev. 7, 1024–1035.
	
	
	\bibitem{MMarte} K. Dholakia, B. W. Drinkwater, M.  Ritsch-Marte 2020 Comparing acoustic and optical forces for biomedical research,  Nature Reviews Physics, 2, 480–491.
	
	\bibitem{CHuertas}   C. S. Huertas, O. Calvo-Lozano  2019 A. Mitchell and Laura M. Lechuga, Advanced Evanescent-Wave Optical Biosensors for the Detection of Nucleic Acids: An Analytic Perspective
	Front. Chem., 25 . 
	
	
	
	\bibitem{LDu} L. Du, A. Yang, A. V. Zayats, X. Yuan 2019
	Deep-subwavelength features of photonic skyrmions in a confined electromagnetic
	field with orbital angular momentum, Nature Physics, 650–654.
	
	\bibitem{Tsesses} S. Tsesses, E. Ostrovsky, K. Cohen, B. Gjonaj, N. H. Lindner and G. Bartal 2018 Optical skyrmion lattice in evanescent electromagnetic fields, 
	Science, 361, 6406, 993-996.
	
	\bibitem{Karen}K. Volke-Sepúlveda, A. O. Santillán, and R. R. Boullosa 2008 Transfer of Angular Momentum to Matter from Acoustical Vortices in Free Space, Phys. Rev. Lett. 100, 024302
	
	\bibitem{Landau} L. D. Landau and E.M. Lifshitz 1987 Fluid Mechanics Oxford: Heinemann.
	
	\bibitem{Bruneau}  	M. Bruneau  2006  Fundamentals of Acoustics London: ISTE Ltd.	
	
	
	\bibitem{Whittaker} E. Whittaker, and G. Watson 1927 A Course of Modern Analysis, Cambridge University Press.
	
	
	\bibitem{Uri} U. Levy, et al 2016 Modes of Free Space, Progress in Optics, 61, 237-281.
	
	\bibitem{Berry}M. V. Berry, Optical currents  2009  J. Opt. A: Pure Appl. Opt. 11, 094001
	
	\bibitem{Yang} Z. Yang 2015 Optical orbital angular momentum of evanescent Bessel waves, Opt. Express 23, 12700-12711.
	
	\bibitem{ZYang1}
	Z. Yang, Xia Zhang, C. Bai, and M. Wang 2018 Nondiffracting light beams carrying fractional orbital angular momentum, J. Opt. Soc. Am. A 35, 452-461.
	
	\bibitem{Gouesbet} G. Gouesbet 2020 Van de Hulst Essay: A review on generalized Lorenz-Mie theories with wow stories and an epistemological discussion, Journal of Quantitative Spectroscopy and Radiative Transfer 253, 107-117.    
	
	\bibitem{Hugo2014} H. E. Hern\'andez-Figueroa, M. Zamboni-Rached and E.  Recami 2013 Non-Diffracting Waves, John Wiley \&  Sons. 
	
	
	\bibitem{OlegV1} O.V. Angelsky, C. Yu Zenkova, P.P. Maksymyak, A.P. Maksymyak, D. I. Ivansky  2019
	Controlling and manipulation of red blood cells by evanescent waves,
	Optica Applicata, Vol. XLIX ,  4.
	
	
	\bibitem{OlegV2}	 O. V. Angelsky, C. Yu Zenkova, S. G. Hanson and Jun Zheng  2020
	Extraordinary Manifestation of Evanescent Wave in Biomedical Application,
	Front. Phys.  https://doi.org/10.3389/fphy.2020.00159
	
	
\end{thebibliography}
\end{document}